\documentclass{aa}
\usepackage{txfonts}
\usepackage{graphicx}

\begin{document}
\title{A fourth component in the young multiple system V 773 Tau
\thanks{Based on observations collected at the German-Spanish Astronomical
 Center on Calar Alto, Spain}}

\author{Jens Woitas}

\institute{Th\"uringer Landessternwarte Tautenburg, Sternwarte 5, 
    07778 Tautenburg, Germany}

\date{Received / Accepted}

\abstract{
 I report on a new component in the pre-main sequence multiple system
 \object{V 773 Tauri}. This second visual companion, V~773~Tau~C,
 with a projected separation of $\approx 0\farcs2$ has been detected
 using speckle interferometry in the near-infrared. Repeated observations
 from 1996 to 2002 show significant orbital motion and thus confirm the
 character of the new companion as a gravitationally bound star.
 Together with the two components of the spectroscopic binary V~773~Tau~A
 and the previously known visual companion V~773~Tau~B, the
 V~773~Tau system appears as a young ``mini-cluster'' of four T~Tauri
 stars within a sphere of a radius less than 100\,AU.
 V~773~Tau~A, B and C form
 a triple system that is not hierarchic, but is apparently stable
 despite of this. The brightness of V~773~Tau~C has probably
 increased over the last years, which may explain its non-detection
 in previous binary surveys.      

}

\maketitle

\section{Introduction}
High angular resolution surveys for companions to T~Tauri stars in nearby star
forming regions (SFRs) and young clusters (cf. Mathieu et al.\,\cite{mat00}
for a review) have revealed that they show a companion frequency at least
as high as among G and K dwarfs in the solar neighbourhood (50 - 60 \%,
e.\,g. Duquennoy \& Mayor \cite{duq91}). This means that multiplicity must be
established already at a very early phase of stellar evolution,
and that binary formation is not an exception but perhaps the dominant
mode of star formation. The probably most surprising outcome of these
binary surveys is that one of the nearest SFRs, the Taurus-Auriga
association, shows a significant binary overabundance compared with
main-sequence dwarfs in the solar neighbourhood and also most other
samples of young stars (K\"ohler \& Leinert \cite{koe98}, and references
therein). If the distribution of orbital periods in Taurus-Auriga is the
same as for main sequence binaries, almost all young stars in this SFR
will be members of binary or multiple systems.
 
Although all known pre-main sequence stars in Taurus-Auriga have been
targets of systematic binary surveys, some companions were apparently
overlooked. Almost all of them are components of triple or higher order
multiple systems (Duch\^{e}ne \cite{duc99},
Richichi et al. \cite{ric99}). In this paper I report on a previously
unknown third companion to \object{V~773~Tau}.

\object{V~773~Tau} is one of the optically brightest members of
the Taurus-Auriga association, and is also a strong radio source.
The latter property allowed a precise determination of its distance,
which is $148\pm 5\,\mathrm{pc}$, using VLBI astrometry
(Lestrade et al.\,\cite{les99}). \object{V~773~Tau} is classified
as a weak-lined T~Tauri star regarding its H$\alpha$
equivalent width of 4\,{\AA} (Herbig \& Bell \cite{hbc88}). 

Leinert et al.\,(\cite{lei93}) and Ghez et al.\,(\cite{ghez93}) have
detected a visual companion to \object{V~773~Tau} with a projected
separation of $\approx 0\farcs15$ using speckle
interferometry in the K-band. Since this companion, hereafter named
\object{V~773~Tau}~B, was close to its periastron at the time of detection,
Tamazian et al.\,(\cite{tam02}) could derive a first orbital solution
based on 9 data points that cover $\approx\,160^{\circ}$ in position angle,
indicating a dynamical system mass of $3.20 \pm 0.71\,M_{\odot}$.
Welty (\cite{wel95}) has shown that the main component, V~773~Tau~A, is
itself a spectroscopic binary with an orbital period of 51 days.
Duch\^{e}ne et al. (\cite{duc01}) have briefly announced the possible
presence of a fourth close component in the \object{V~773~Tau} system. In this
paper astrometric and photometric properties of this object are given,
and its physical membership to \object{V~773~Tau} is confirmed from its
orbital motion. 

\begin{figure*}
\centering
\includegraphics[width=6.5cm]{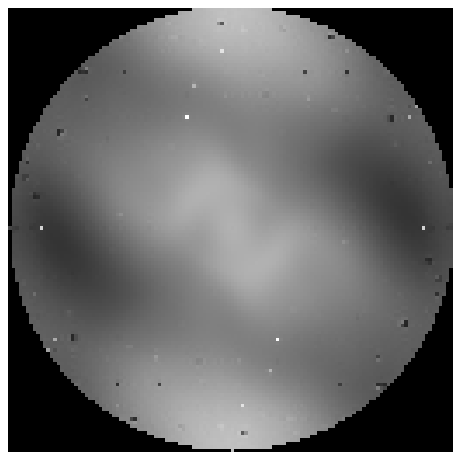}
\includegraphics[width=6.5cm]{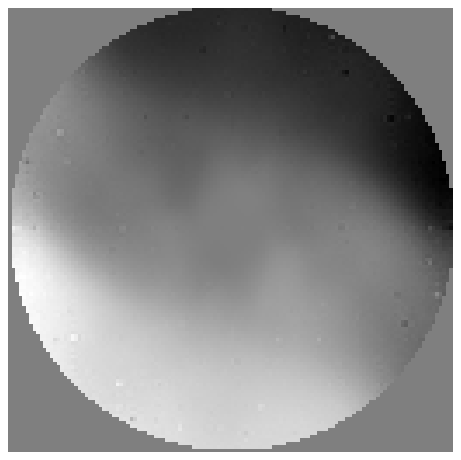}
\vspace{0.2cm}
\includegraphics[width=6.5cm]{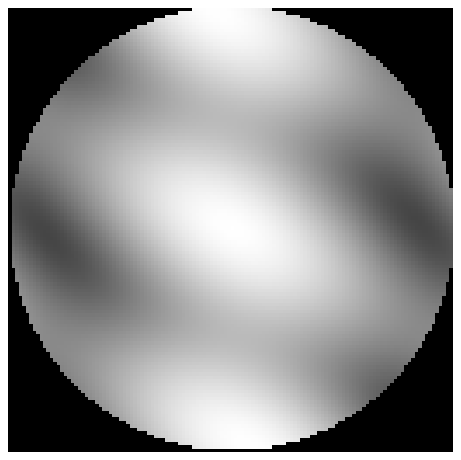}
\includegraphics[width=6.5cm]{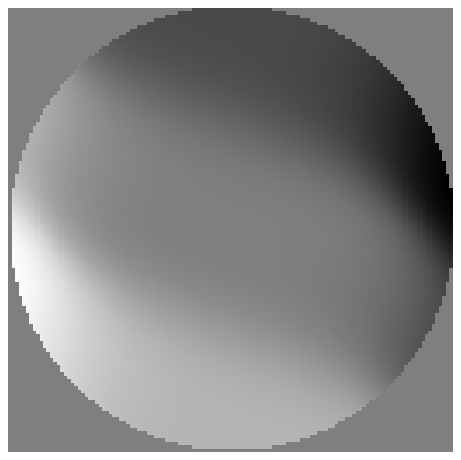}
\includegraphics[width=6.5cm]{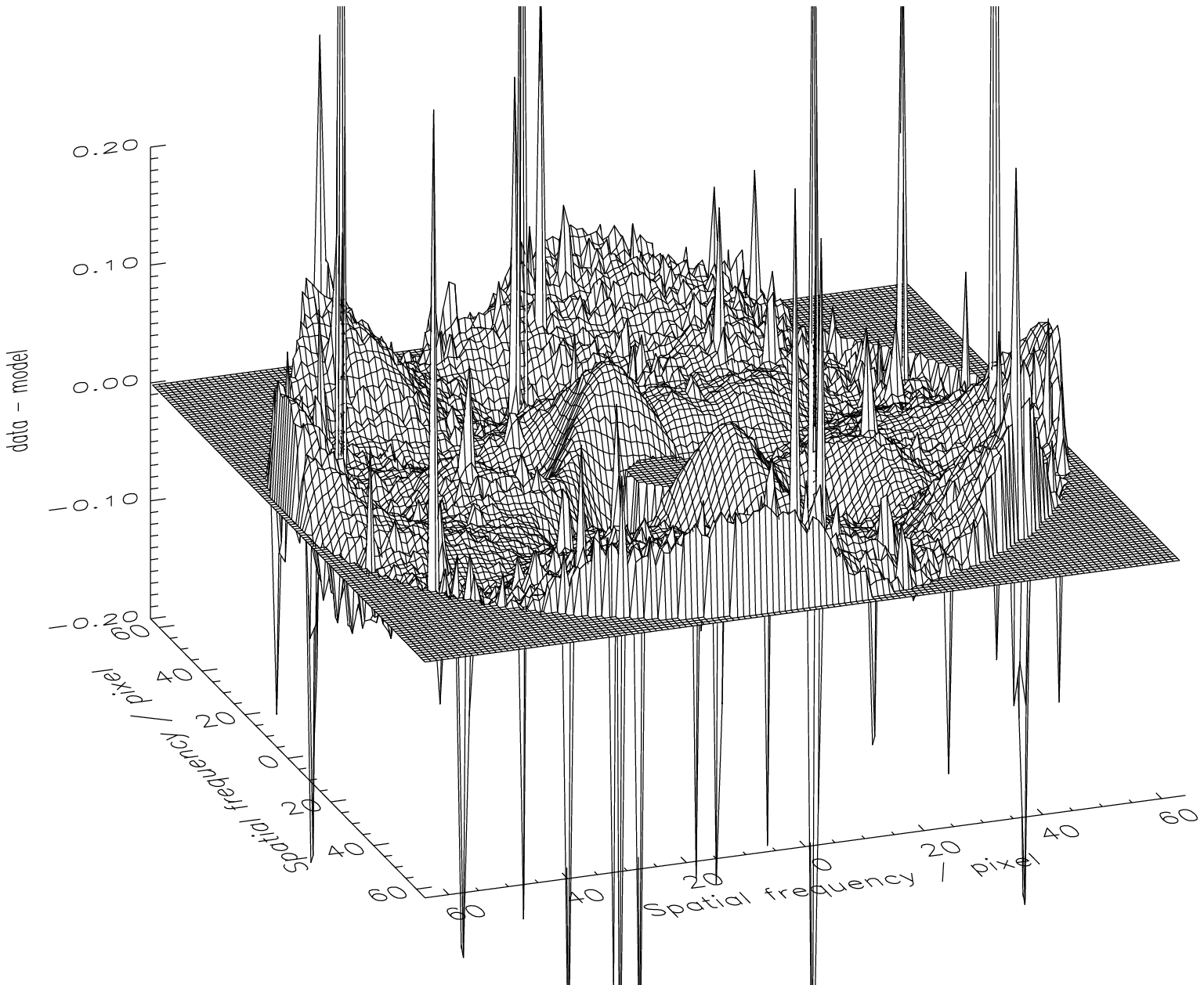}
\includegraphics[width=6.5cm]{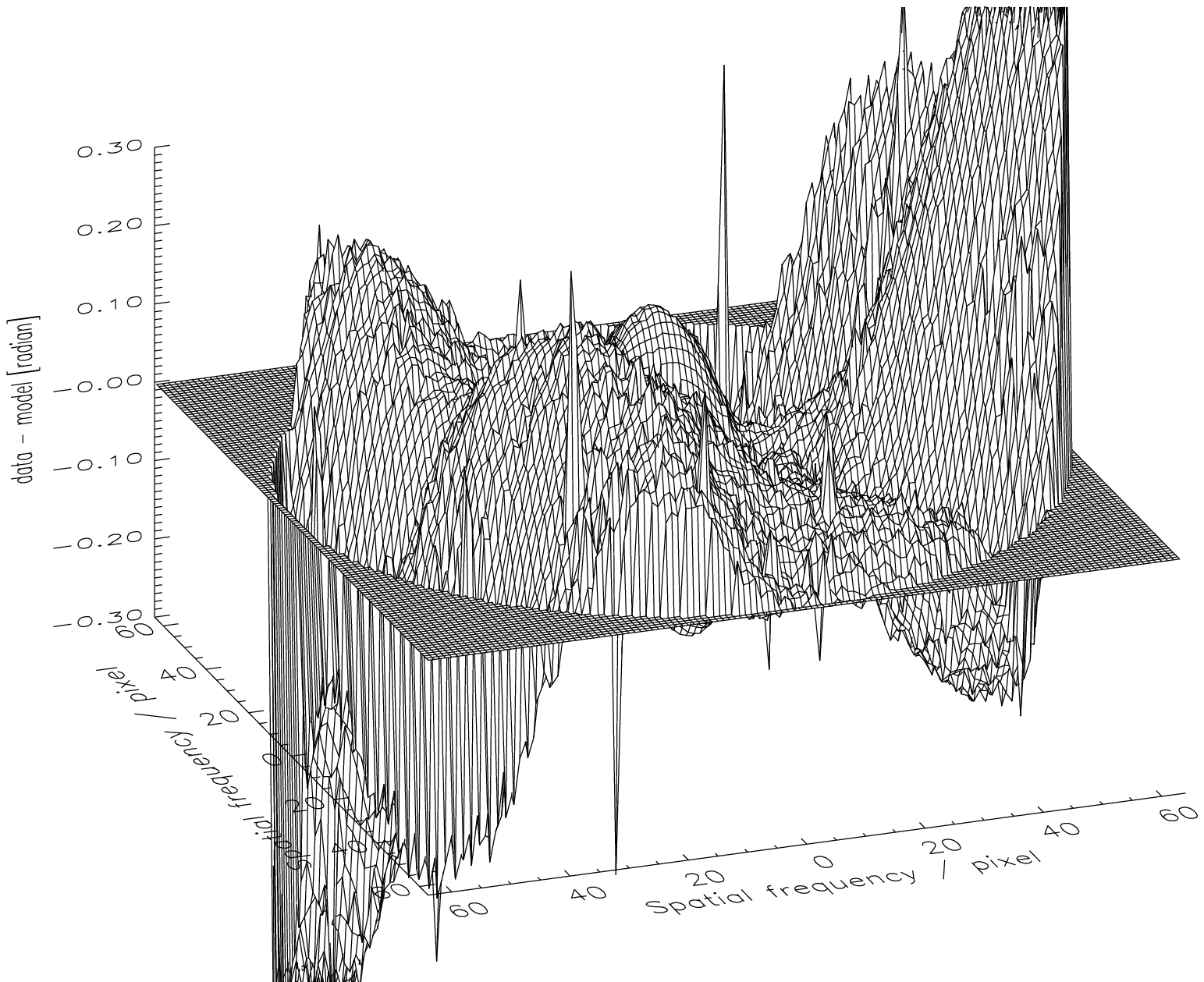}
\caption{\label{vispha} First line: modulus (left) and bispectrum phase
 (right) of the complex visibility for V~773~Tau, derived from 3000 short
 ($\tau = 0.13\,\mathrm{s}$) exposures obtained in the K-band with the
 Omega Cass camera at the 3.5\,m-telescope at Calar Alto on 19 October 2002.
 The images have been cut off at the Nyquist spatial frequency that is
 5.3\,arcsec$^{-1}$ for the adopted pixel scale of 0\farcs095/pixel.
 North is up, and east is to the left.
 Second line: Artificial modulus (left) and phase (right) that are
 constructed from a triple star model and fit to the data. Last line:
 residuals of the fit. The modulus is normalized to one, the phase
 is given in radian.}
\end{figure*}

\section{Observations and data analysis}
\label{obs}
\object{V~773~Tau} has been observed with the near-infrared camera
Omega Cass at the 3.5\,m-telescope on Calar Alto in February 2001,
November 2001 and October 2002. All observations were carried out
in the K-band at $\lambda = 2.2\,\mu$m. Omega Cass is in its subarray
mode capable of taking fast sequences of short exposures with integration
times $\tau\approx 0.1\,\mathrm{s}$. This allows the application
of speckle interferometry as a high angular resolution technique
that yields information about object structures down to the diffraction
limit that is $\lambda/D\approx 0\farcs13$ for the given instrumentation.
If it is known from previous observations that the object is a binary,
then its parameters can be reliably determined even down to a projected
separation of $\lambda/(2D)$.
In each observing run at least 1500 short exposures were taken for
V~773~Tau as well as for the nearby single star \object{SAO~76511} that is 
used as PSF calibrator. These exposures are stored as ``data cubes'' of 250
frames. After each data cube the telescope position is switched between
the object and the PSF calibrator. This ensures that both are observed
at nearly the same airmass and seeing conditions. The mean power spectrum
of one data cube for the object is deconvolved with that for the
reference star. The phase is reconstructed using the algorithm by
Knox \& Thompson (\cite{kt74}) and the bispectrum method
(Lohmann et al.\,\cite{loh83}). Finally, modulus and phase of the
complex visibility are averaged over all data cubes. These mean
complex visibilities, derived from all observations of
V~773~Tau within one night, are used to determine the parameters
of the visual companions, i.\,e. their position angles, projected
separations and flux ratios with respect to the primary. For this
purpose artificial complex visibilities are constructed from
sets of triple star models and fit to the data.
\footnote{A more detailed description of the data reduction and
analysis process can be found in K\"ohler et al. (\cite{koe00}).}
As an example data, model and the residuals of the fit are shown for the
observation at 19 Oct 2002 in Fig.\,\ref{vispha}. The stripe patterns
(from upper right to lower left in the images) are the typical sign
of a stellar companion in Fourier space. Note also the shading at the
right and left hand sides of the images. This is caused by the presence of
\object{V~773~Tau}~B that is resolved although its projected
separation with respect to \object{V~773~Tau}~A is only 0\farcs115.

The pixel scale is  $\approx 0\farcs095/\mathrm{pixel}$ for
Omega Cass and was, as well as the detector orientation, empirically
determined for the individual observations using astrometric fits
to images of the Orion Trapezium cluster core where precise astrometry
has been given by McCaughrean \& Stauffer (\cite{mcc94}).

In analyzing the data one has to consider that the Nyquist frequency,
defined as half the sampling frequency

\begin{equation}
f_N = f_s/2 = \frac{1}{2\cdot 0\farcs095} = 5.3\,\mathrm{arcsec}^{-1}
\end{equation}

is less than the cutoff frequency

\begin{equation}
f_c = \frac{D}{\lambda} = \frac{1}{0\farcs13} = 7.7\,\mathrm{arcsec}^{-1} 
\end{equation}

for the instrument and telescope configuration used. This may cause
aliasing effects for spatial frequencies larger than 

\begin{equation}
f_s - f_c = 2.9\,\mathrm{arcsec}^{-1}.
\end{equation}

In Fig. \ref{vispha} this corresponds to a radius of R~=~35~pixel around
the images centres, while the whole data range used for the
fit is up to R~=~64 pixel, i.\,e. the Nyquist spatial frequency.
This has probably not a large influence on the general picture.
V~773~Tau~C will be resolved already at $f \le 35\,\mathrm{pixel}$,
and the relative position of V~773~Tau~B matches well the orbital
fit by Tamazian et al.\,(\cite{tam02}) and thus seems reliable.
For spatial frequencies $f > 35\,\mathrm{pixel}$ the signal is however
affected by an unknown bias. Therefore any results about flux ratios and
photometry for very close companions like V~773~Tau~B and C are highly
uncertain for Omega Cass data.

\begin{table*}
\caption{\label{obs-data} Relative astrometry and flux ratio of
 the new companion V~773~Tau~C with respect to V~773~Tau~Aa obtained
 from speckle interferometric observations at the 3.5\,m-telescope
 on Calar Alto. For the last observations from Feb 2001 to Oct 2002
 also the results for component V~773~Tau~B are given.}  
\begin{tabular}{llllllllll}
Date & Instrument & Companion & Position angle [deg] & Separation$[``]$ &
 Flux ratio & $\chi^2_{modulus}$ & $\chi^2_{phase}$  & $\chi^2_{modulus}$ &
 $\chi^2_{phase}$ \\
   & & & (from North to East) & & ($F_{comp}/F_A$ &
 \multicolumn{2}{c}{(uncorrected)} & \multicolumn{2}{c}{(corrected)} \\
   & &             &         & & at $\lambda = 2.2\mu\mathrm{m}$) \\
  \hline
27 Sep 1996 & MAGIC      & C & 174 $\pm$ 5  & 0\farcs226 $\pm$ 0\farcs015
  &  $\le 0.09$\footnotemark  & 0.012 & 0.334 & 0.96 & 26.8 \\
9 Feb 2001  & Omega Cass & C & 160 $\pm$ 10 & 0\farcs223 $\pm$ 0\farcs020
  & 0.12 $\pm$ 0.02  & 0.996 & 0.211 & 65.4 & 14.5 \\
            &            & B & 93.5 $\pm$ 0.6 & 0\farcs095 $\pm$ 0\farcs005
  & 0.41 $\pm$ 0.02  \\
3 Nov 2001  & Omega Cass & C & 154 $\pm$ 10 & 0\farcs203 $\pm$ 0\farcs020
  & 0.15 $\pm$ 0.02  & 0.459 & 0.068 & 17.0 & 2.52 \\
            &            & B & 94.6 $\pm$ 1.0 & 0\farcs099 $\pm$ 0\farcs007
 & 0.49 $\pm$ 0.02 \\
19 Oct 2002 & Omega Cass & C & 155.7 $\pm$ 2.0 & 0\farcs232 $\pm$ 0\farcs010
 & 0.21  $\pm$ 0.02  & 0.002 & 0.111 & 0.46 & 25.6 \\
            &            & B &  97.4 $\pm$ 1.0 & 0\farcs115 $\pm$ 0\farcs010
 & 0.36 $\pm$ 0.01 \\
 \hline
\end{tabular}
\end{table*}

\footnotetext{V~773~Tau~A and B were not resolved in this observation.
 The measured flux ratio is $F_C/F_{A+B} = 0.06\pm 0.01$.}

After detecting the new companion in the Omega Cass data, I have
re-analysed near infrared speckle interferometric data of
\object{V~773~Tau} obtained with the MAGIC camera also at the
3.5\,m-telescope on Calar Alto in September 1996. The pixel scale
of this camera is $0\farcs07/\mathrm{pixel}$, corresponding to
a Nyquist frequency of $7.1\,\mathrm{arcsec}^{-1}$. This means that
the mentioned aliasing problem does barely exist here, and the
derived binary parameters are thus almost unbiased by this effect.
Evidence for the presence of \object{V~773~Tau}~C could indeed be found.
At the time of this observation the brighter companion \object{V~773~Tau}~B
had a projected separation of less than 0\farcs05 from the primary
(Tamazian et al.\,\cite{tam02}). The pair V~773~Tau~AB thus
appears as a point source even in a diffraction limited image, and the
contribution of V~773~Tau~B to the complex visibility is approximately
constant for all power spectrum frequencies. The flux ratio has
therefore to be read as $F_C/F_{A+B} = 0.06\pm 0.01$ for this observation.
This is at the edge of the dynamical range accessible with speckle
methods, but as only a double star model has to be used here, 
the parameters of V~773~Tau~C can be determined from these data with
sufficient accuracy.

To test the reliability of the fits I have calculated the $\chi^2$,
defined here as

\begin{equation}
\label{chi}
\chi^2 = \frac{1}{n-1} \sum_{k=1}^{n} \frac{(x_{data}(k) - x_{model}(k))^2}{\sigma^2(k)},
\end{equation}

where the sum is over all pixels used for the fit. $x$ and $\sigma$
are the values of modulus or phase and, respectively, the standard deviations
for individual pixels resulting from averaging over all
data cubes. While this $\chi^2$ is a reasonable indicator for
the quality of the fit, its mathematical interpretation should be
taken with care. The number of degrees of freedom is probably not
$n - 1$ because the pixels are correlated over a distance (in frequency
space) that is defined by the reciprocal seeing width. This means
that the $\chi^2$ values have to be multiplied with a factor that
can be estimated as

\begin{equation}
\label{correction}
C = \frac{FOW}{(FWHM_{seeing})^2}.
\end{equation}

The field of view (FOV) is $8\farcs96^2$ for MAGIC and
$12\farcs16^2$ for the subarray of Omega Cass. In Table \ref{obs-data}
an overview of all observations of V~773~Tau~C is presented,
and the corrected and uncorrected $\chi^2$ values
are given. The latter ones indicate that
for all observations the error of the fit is on average smaller
than the 1$\sigma$ uncertainty of the data.

 Relative astrometry and
photometry is also given for \object{V~773~Tau}~B for the
last three observations. A complete list of all binary parameter
measurements of this latter companion since 1990 can be found in Tamazian 
et al.\,(\cite{tam02}, Table 1 therein).

\begin{figure}
\resizebox{\hsize}{!}{\includegraphics{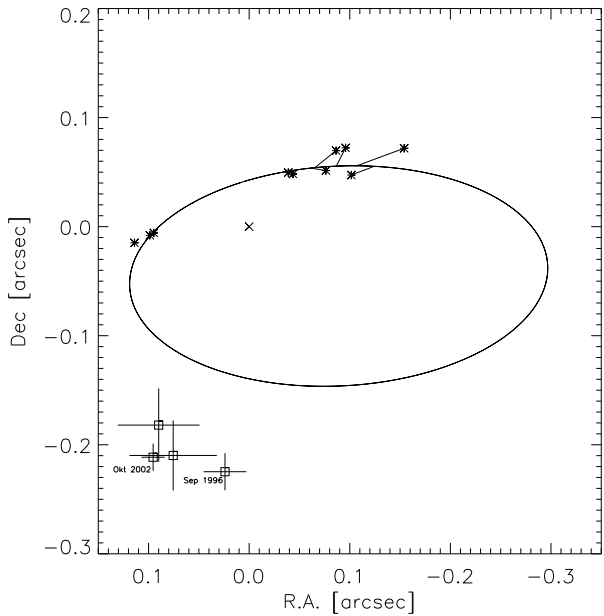}}
 \caption{\label{orbit} Orbital motion in the V~773~Tau system.
  The cross at (0.0, 0.0) marks the primary V~773~Tau~A.
  The asterisks denote the relative positions of V~773~Tau~B together
  with the orbital fit by Tamazian et al.\,(\cite{tam02}). The squares
  represent relative position measurements of the new companion
  V~773~Tau~C from September 1996 (most right) to October 2002 (most left)}
\end{figure}

\section{Discussion}

\subsection{Reliability of the detection}
Since the projected separations of V~773~Tau~B measured in and after
February 2001 are in the order of
both the diffraction limit and the pixel scale, one may doubt that
\object{V~773~Tau} really is a visual triple star. If this were the case,
the ``new'' companion would just be \object{V~773~Tau}~B that has appeared
on the opposite side of the primary after passing its periastron.
There are several arguments against this idea: The signature of
\object{V~773~Tau}~B is seen in Fig. \ref{vispha}  
at a position different from \object{V~773~Tau}~C, and assuming that
\object{V~773~Tau} is a visual triple star significantly improves the
model fit mentioned in Sect.\,\ref{obs}. For the data from October 2002
presented in Fig.\,\ref{vispha}, a fit only with a double star model
(i.\,e. without V~773~B) yields a (corrected) $\chi^2 = 4.85$ for the modulus
and $\chi^2 = 112.0$ for the phase. In particular for the modulus
this is about ten times larger than the value given in Table\,\ref{obs-data}.
The triple star solution has been reproduced in three data sets
obtained at different epochs. Furthermore, the data points obtained
for \object{V~773~Tau}~B until 1994 and for the new companion after
1996 cannot be explained by orbital motion of {\it one} object since
such a companion would have passed through its periastron much too fast.

Another question is if  \object{V~773~Tau}~C really is a gravitationally
bound stellar companion. To check this, one has to prove if its
relative motion with respect to the primary is orbital motion.
A method for this has been discussed in detail in a previous paper
(Woitas et al.\,\cite{woi01}). Briefly, from a linear fit to the
four position measurements of \object{V~773~Tau}~C from 1996 to 2002,
I derive a tangential velocity of $v_t = 8.7 \pm 2.6\,\mathrm{km/s}$.
This means that the relative motion is significant on the 3$\sigma$
level and that modulus and direction of this motion are very well
explainable with orbital motion. A background star projected by chance
close to \object{V~773~Tau}~A would move in a direction antiparallel
to the proper motion of \object{V~773~Tau} ($\mu_{\alpha}$ = 0.65 mas/yr,
$\mu_{\delta}$ = -24.89 mas/yr from the {\it Hipparcos} catalogue).
This translates into a relative motion of a projected companion of
17.5 km/s in northern direction which is inconsistent with the data.
Furthermore, one can estimate from the low stellar density in
Taurus-Auriga that the probability to find a chance-projected object
brighter than K~=~12\,mag within a radius of 0\farcs3 around an
association member is only $1.1\cdot 10^{-5}$ (Leinert et al.\,\cite{lei93}).
Although I cannot give absolute K-band photometry for \object{V~773~Tau}~C,
it is certainly brighter than K~=~12\,mag, given the mean system magnitude
$K = 6.45 \pm 0.09\,\mathrm{mag}$ (Kenyon \& Hartmann \cite{ken95})
and the flux ratios from Table\,\ref{obs-data}.
Any random projections are thus very unlikely. As result
of this discussion I conclude that there are really two visual companions
in the \object{V~773~Tau} system, which makes this system a quadruple,
in combination with the close spectroscopic companion to 
\object{V~773~Tau}~A detected by Welty (\cite{wel95}).
The fact that V~773~Tau~C has not been detected in previous
observations might be the outcome of a long term variability
of its brightness. As discussed in Sect.\,\ref{obs} the flux ratios
$F_B/F_A$ and $F_C/F_A$ derived from Omega Cass data are affected 
with a bias resulting from aliasing at higher spatial frequencies
and have therefore to be taken with caution.
The value $F_C/F_{A+B} = 0.06$ obtained with MAGIC in 1996 is however
not affected with this bias. If one assumes $F_B/F_A \le 0.5$, which
seems justified looking at Table\,\ref{obs-data} and also at the
(unbiased) value of $F_B/F_A = 0.13\pm 0.04$ given by Leinert et al.
(\cite{lei93}), one ends up with $F_C/F_A \le 0.09$ for this observation.
On the other hand, the value for the
observation in Oct 2002 is indeed around  $F_C/F_A \approx 0.2$. This
can be proven by lowering the fit radius to $f = 35\,\mathrm{pixel}$
where the bias disappears. Since V~773~Tau is a weak-lined
T~Tauri star with no signs of active circumstellar accretion,
variable extinction might explain this unexpected strong
variability.

\subsection{The V 773 Tauri ``mini-cluster''}
The real (i.\,e. de-projected) separation between \object{V~773~Tau}~C
and \object{V~773~Tau}~A cannot be determined from my data. The
presence of measurable orbital motion however indicates that this separation
is much less than 100\,AU. For this semimajor axis the annual orbital motion
would be $0.6^{\circ}/\mathrm{yr}$ in position angle for the idealized
situation of a circular orbit viewed pole-on and a system mass of
$3\,M_{\odot}$. The latter is similar to the dynamical system mass derived
for V~773~Tau~AB by Tamazian et al.\,(\cite{tam02}). The observed orbital
motion is much faster, so 100\,AU is a conservative upper limit for the
separation between \object{V~773~Tau}~A and C. The semimajor axis of component
\object{V~773~Tau}~B is $a_{AB} \approx 37\,\mathrm{AU}$ (Tamazian
et al.\,\cite{tam02}). This means that together with the spectroscopic
binary \object{V~773~Tau}~A, the whole system consists of four stars within
a sphere with a radius that is less than 100\,AU. To my
knowledge, no other similar ``mini-cluster''
of T~Tauri stars is known at this time. The other young quadruple systems
in Taurus-Auriga (\object{GG~Tau}, \object{UZ~Tau}, \object{UX~Tau}
and \object{BD+26\,718B}) consist of components separated by several
arcseconds. The very small size of the \object{V~773~Tau} system
strongly indicates that all components have formed from the same
molecular cloud core and thus must be coeval to a high degree.
This makes them well suited for an empirical test of theoretical isochrones
like in the study that White et al. (\cite{whi99}) have done
for the young quadruple system \object{GG~Tau}.  

The semimajor axes of \object{V~773~Tau}~AB and AC have probably the
same order of magnitude, which means that \object{V~773~Tau}~ABC is
not a hierarchical triple system. This poses the question if the whole
system is dynamically stable. This is probably the case. Although
age determinations from pre-main sequence evolutionary models are
highly uncertain, one can fairly assume that \object{V~773~Tau}
is at least some $10^5\,\mathrm{yr}$ old since it is a Class\,III
source with no signs of infalling envelopes or strong circumstellar
accretion. This age is much larger than the orbital periods of
the companions (125\,yr for V~773~Tau~AB, Tamazian et al.\,\cite{tam02}),
so if there were dynamical instabilities they would
have already destroyed the system. It remains however unclear why
this is not the case. Another striking property of the V~773~Tau
system seen in Fig.\,\ref{orbit} is that both visual companions
apparently move in different directions on the sky. This means that
they either do not orbit in the same direction or have quite different orbital
inclinations. For a better understanding of this situation
additional observations will be needed to include \object{V~773~Tau}~C
into the orbit calculation for \object{V~773~Tau}~AB by
Tamazian et al.\,(\cite{tam02}) and then derive the orbital parameters
of the whole triple system.

The detection of only one additional companion has of course no
significant impact on binary statistics. The finding of an additional
component in a very well studied young multiple system like
\object{V~773~Tau} however indicates that there may be even more
companion stars in Taurus-Auriga and that higher order multiple systems
may be more common among pre-main sequence stars than
was previously thought. The known binary stars in Taurus-Auriga
and other SFRs are for this reason interesting targets for
next-generation high angular resolution instruments that will
be able to close the detection gap between spectroscopic and
visual companions.

\begin{acknowledgement}
I acknowledge a thoughtful and constructive report from the
anonymous referee that led to a significant improvement of
this paper. J.W. is supported from the Deutsches Zentrum
f\"ur Luft- und Raumfahrt under grant number 50 OR 0009.
The observations on Calar Alto were supported by travel money from
the Deutsche Forschungsgemeinschaft (grant numbers Wo~834~/~1-1,
Wo~834~/~2-1 and Wo~834~/~4-1).
The data reduction has been performed using the ``Binary/Speckle'' package
developed by Rainer K\"ohler. I am also grateful to the night assistants
and the technical staff on Calar Alto for their support during my
observing runs.   
\end{acknowledgement}

\end{document}